# Brain Damage and Motor Cortex Impairment in Chronic Obstructive Pulmonary Disease: Implication of Nonrapid Eye Movement Sleep Desaturation

Francois Alexandre, PhD[1,2]; Nelly Heraud, PhD[2,3]; Anthony M.J. Sanchez, PhD[4,5]; Emilie Tremey, PhD[2,3]; Nicolas Oliver, MD[2]; Philippe Guerin, MD[3]; Alain Varray, PhD[1]

[1]Movement To Health Laboratory, Euromov, University of Montpellier, Montpellier, France; [2]Clinique du Souffle La Vallonie, Fontalvie, Lodève, France; [3]Clinique du Souffle Les Clarines, Fontalvie, Riom-es-Montagnes, France; [4]UMR866 Dynamique Musculaire et Métabolisme, INRA, University of Montpellier, Montpellier, France; [5]Laboratoire Performance Santé Altitude, EA 4604, University of Perpignan Via Domitia, Font-Romeu, France

**Study Objectives:** Nonrapid eye movement (NREM) sleep desaturation may cause neuronal damage due to the withdrawal of cerebrovascular reactivity. The current study (1) assessed the prevalence of NREM sleep desaturation in nonhypoxemic patients with chronic obstructive pulmonary disease (COPD) and (2) compared a biological marker of cerebral lesion and neuromuscular function in patients with and without NREM sleep desaturation.
**Methods:** One hundred fifteen patients with COPD (Global Initiative for Chronic Obstructive Lung Disease [GOLD] grades 2 and 3), resting $PaO_2$ of 60–80 mmHg, aged between 40 and 80 y, and without sleep apnea (apnea-hypopnea index < 15) had polysomnographic sleep recordings. In addition, twenty-nine patients (substudy) were assessed i) for brain impairment by serum S100B (biological marker of cerebral lesion), and ii) for neuromuscular function via motor cortex activation and excitability and maximal voluntary quadriceps strength measurement.
**Results:** A total of 51.3% patients (n = 59) had NREM sleep desaturation ($NREM_{Des}$). Serum S100B was higher in the $NREM_{Des}$ patients of the substudy (n = 14): 45.1 [Q1: 37.7, Q3: 62.8] versus 32.9 [Q1: 25.7, Q3: 39.5] pg.ml$^{-1}$ (P = 0.028). Motor cortex activation and excitability were lower in $NREM_{Des}$ patients (both P = 0.03), but muscle strength was comparable between groups (P = 0.58).
**Conclusions:** Over half the nonhypoxemic COPD patients exhibited NREM sleep desaturation associated with higher values of the cerebral lesion biomarker and lower neural drive reaching the quadriceps during maximal voluntary contraction. The lack of muscle strength differences between groups suggests a compensatory mechanism(s). Altogether, the results are consistent with an involvement of NREM sleep desaturation in COPD brain impairment.
**Clinical Trial Registration:** The study was registered at www.clinicaltrials.gov as NCT01679782.
**Citation:** Alexandre F, Heraud N, Sanchez AM, Tremey E, Oliver N, Guerin P, Varray A. Brain damage and motor cortex impairment in chronic obstructive pulmonary disease: implication of nonrapid eye movement sleep desaturation. *SLEEP* 2016;39(2):327–335.

### Significance

This study reveals that over half of COPD patients (grade 2 and 3) nonhypoxemic at rest have nocturnal desaturation during nonrapid eye movement sleep stages. These patients also present an increased cerebral lesion biomarker and a reduced motor cortex activation and excitability during quadriceps voluntary contractions. These results are consistent with the development of cerebral lesions in case of nonrapid eye movement sleep desaturation, and corroborate the hypothesis of an absence of cerebrovascular reactivity during these stages. The prevention of nonrapid eye movement sleep desaturation thus appears as a relevant clinical perspective to prevent COPD brain injury or even to restore brain function.

## INTRODUCTION

Patients with chronic obstructive pulmonary disease (COPD) present several neurological disorders that directly affect daily life. These disorders include cognitive dysfunction, which degrades quality of life by, for example, decreasing driving ability.[1] In our laboratory, we previously showed that motor cortex impairment could be involved in COPD muscle weakness due to inadequate motor cortex activation.[2]

The origin of the cerebral dysfunction in patients with COPD remains unelucidated. The potential role of hypoxemia in triggering neuronal damage and dysfunction by cerebral oxygen deprivation has often been hypothesized.[3] However, several studies have provided evidence of cerebral dysfunction in nonhypoxemic COPD patients, indicating that hypoxemia *per se* is not the main factor.[4–6] This observation is unsurprising because an adequate oxygen supply to the brain is permanently ensured through cerebrovascular oxygen ($O_2$) reactivity. During hypoxemia or oxygen desaturation, cerebrovascular $O_2$ reactivity prevents cerebral hypoxia by increasing cerebral blood flow (CBF) up to 200%.[7,8] Consequently, the resting CBF is much higher in hypoxemic than in nonhypoxemic COPD patients and healthy controls.[9,10] For the same reason, CBF increases in COPD during exercise-induced desaturation.[11] This results in adequate cerebral oxygen delivery even in the case of hypoxemia.[11] As a whole, these studies provide evidence that cerebrovascular $O_2$ reactivity prevents brain hypoxia in COPD.

Unfortunately, cerebrovascular reactivity is impaired during nonrapid eye movement (NREM) sleep stages.[12–15] Numerous studies have reported an unexpected absence of CBF modulation during NREM sleep (but not during rapid eye movement [REM] sleep) in individuals who experience NREM sleep desaturation.[12–15] Indeed, by decreasing arterial saturation of oxygen ($SaO_2$) artificially by 5% to 10%, Meadows et al.[15] found a consistent CBF increase in wake states, whereas it tended to decrease during slow wave sleep in hypoxemia. Therefore, if the arterial oxygen content falls below the normal value during NREM sleep, it may not be compensated, potentially leading to neuronal injury.[16]

Nocturnal desaturation is frequent in nonhypoxemic COPD patients. The prevalence of COPD patients who are normoxic while awake and who spend at least 30% of the total sleep time (TST) with a saturation of peripheral oxygen ($SpO_2$) below 90% ranges from 38% to 70%.[17–19] To the best of our knowledge, the prevalence of NREM sleep desaturation in COPD has never been specifically assessed. It is generally acknowledged that the deepest desaturation occurs during REM sleep. However, because REM sleep represents only about 13% of the TST in COPD,[20] patients with COPD and nocturnal desaturation (for at least 30% of TST) also necessarily experience desaturation for a significant proportion of NREM sleep.

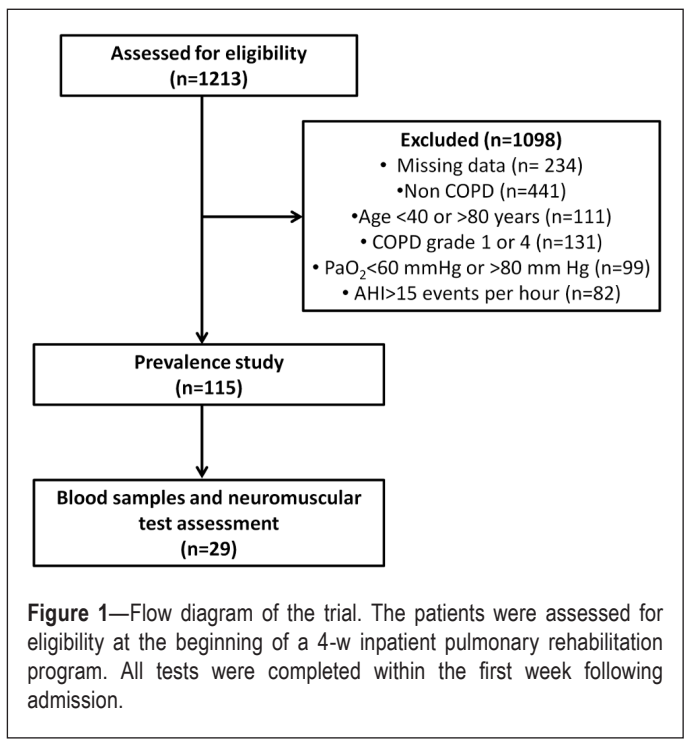

**Figure 1**—Flow diagram of the trial. The patients were assessed for eligibility at the beginning of a 4-w inpatient pulmonary rehabilitation program. All tests were completed within the first week following admission.

Central nervous system (CNS) injury in COPD was recently evidenced by magnetic resonance imaging (MRI) and measurement of serum S100B levels.[21,22] S100B is a calcium binding protein, mainly produced by astrocytes,[23] that is released in the blood circulation in response to glia cell activation during acute and chronic conditions of brain damage.[24] An increase in serum S100B concentration has been described in a wide range of neurological disorders such as acute ischemic and traumatic brain injury and hypoxic brain damage.[25–27] Serum S100B is considered as a surrogate biomarker for neuronal injury[28] and has the main advantage of providing an easy-to-use assessment of cerebral damage.[29]

The aim of the study was twofold: to determine the prevalence of patients with COPD who are nonhypoxemic but experience nocturnal desaturation during NREM sleep; and to compare CNS injury and neuromuscular function in patients experiencing desaturation and those who are not during NREM sleep and assess the repercussions of desaturation on neural drive during maximal voluntary muscle contraction. We hypothesized higher levels of serum S100B associated with lower motor cortex activation and lower muscle strength in patients with COPD who experience desaturation during NREM sleep.

## METHODS

### Participants

The study was conducted between 2012 and 2014 at the Clinique du Souffle La Vallonie in Lodeve, France, and the Clinique du Souffle Les Clarines in Riom-es-Montagnes, France.

Over this period, 1,213 patients taking part in a 4-w inpatient pulmonary rehabilitation program underwent a routine medical examination in the first days following admission, composed of anthropometric evaluation, resting pulmonary function assessment, resting blood gas assessment, the 6-min walk test, and polysomnographic sleep (PSG) recordings. after completion, patient records were screened to identify those patients who met the following criteria: between 40 and 80 y old, diagnosis of COPD with postbronchodilator forced expiratory volume in 1 sec ($FEV_1$) between 30% and 80% of predicted values (corresponding to grades 2 and 3 of the Global Initiative for Chronic Obstructive Lung Disease [GOLD] classification[30]), resting partial pressure of oxygen ($PaO_2$) between 60 and 80 mmHg, and an apnea-hypopnea index (AHI) lower than 15 events per hour. One hundred fifteen patients fulfilled these criteria and were thus selected for a study to determine the prevalence of NREM sleep desaturation in nonhypoxemic COPD patients (Figure 1).

In a second step, we compared the neuromuscular function between patients with COPD with and without NREM sleep desaturation. Over a 6-mo period, a total of 29 consecutive patients underwent additional blood sampling and neuromuscular assessment. Patients were not eligible for neuromuscular assessment if they were unable to give written consent or perform the experimental maneuvers, were on medication known to impair brain function, or had impaired visual function, a pacemaker, current or past alcohol abuse, an exacerbation in the past 4 w, or neurologic or neuromuscular disease. Procedures were approved by the local Ethics Committee (Comité de protection des personnes Sud Est VI, number AU980) and complied with the principles of the Declaration of Helsinki for human experimentation. The study was registered at www.clinicaltrials.gov as NCT01679782.

### Design

All tests were completed within the first week after admission. All participants were first evaluated for anthropometric parameters, resting pulmonary function, resting blood gases, the 6-min walk test, and polysomnographic sleep (PSG) recordings. The patients eligible for neuromuscular assessment were then probed and underwent medical examination after giving written consent. These patients were familiarized with the neuromuscular tests on the first day. Blood samples were collected the next day at patient wake-up, and between 06:30 and 07:30. The neuromuscular tests took place in the morning. The design of the neuromuscular tests is detailed in the Protocol section.

### Measurements

#### Pulmonary function test
Diagnosis and staging of COPD were based on spirometry (V6200 Autobox, Sensormedics Corp., Yorba Linda, CA, USA). Measurements included forced vital capacity (FVC) and $FEV_1$. The presence of persistent airflow obstruction and thus COPD was defined by a postbronchodilator $FEV_1$/FVC ratio < 70%. The $FEV_1$ values were compared with the predicted values of Quanjer et al.[31]

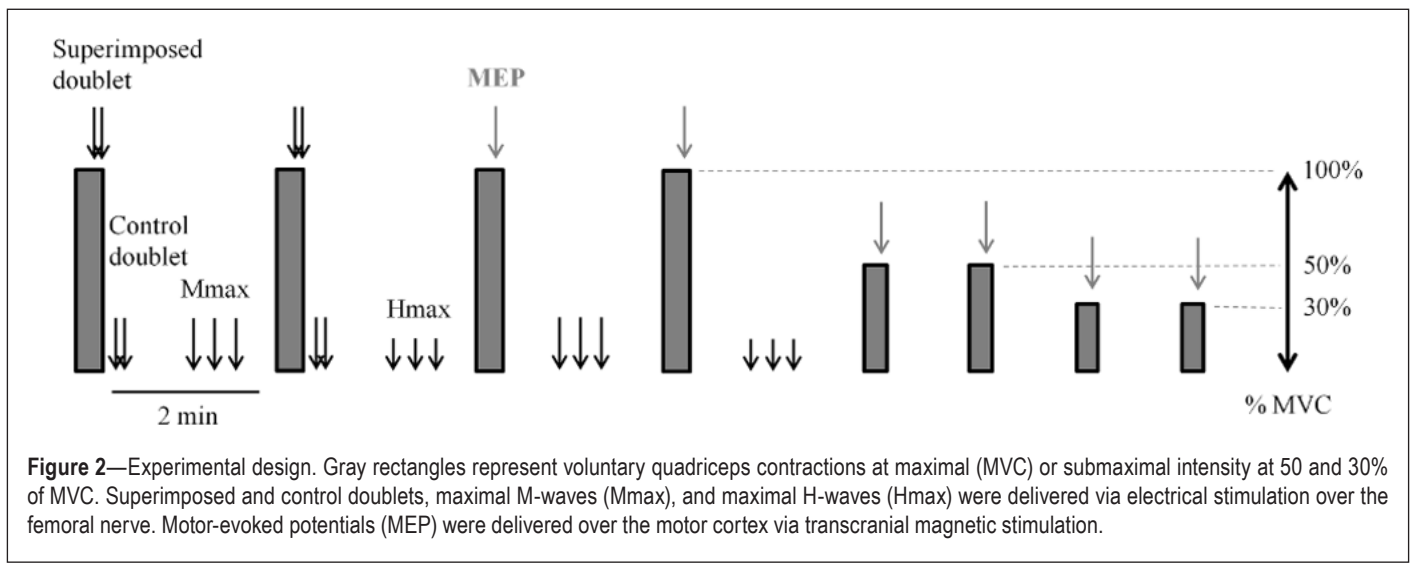

Figure 2—Experimental design. Gray rectangles represent voluntary quadriceps contractions at maximal (MVC) or submaximal intensity at 50 and 30% of MVC. Superimposed and control doublets, maximal M-waves (Mmax), and maximal H-waves (Hmax) were delivered via electrical stimulation over the femoral nerve. Motor-evoked potentials (MEP) were delivered over the motor cortex via transcranial magnetic stimulation.

### Blood gas analysis

Blood gases ($PaO_2$ and partial arterial pressure of carbon dioxide [$PaCO_2$]) collected from the radial artery were measured in the resting patients while they breathed room air, using a blood gas analyzer (ABL 825, Radiometer Medical, Bronshoj, Denmark).

### Polysomnographic sleep recordings

PSG was performed using standard techniques and manually analyzed according to the latest guidelines of the American Academy of Sleep Medicine.[32] Stage epoch classification and $SpO_2$ were exported at 1 Hz in a text file. Then the percentage of $SpO_2$ below 90% during NREM sleep was analyzed with an automatic routine developed in MATLAB (MATLAB 8.0, The MathWorks, Inc., Natick, MA, USA). Patients who spent more than 10% with $SpO_2$ below 90% during NREM sleep were classified as NREM sleep desaturators ($NREM_{Des}$), and non-NREM sleep desaturators ($NREM_{noDes}$) otherwise, according to published data giving evidence of cognitive dysfunction for similar levels of desaturation.[33]

### Exercise-induced desaturation

Exercise-induced desaturation was assessed during a 6-min walk test,[34] which was performed indoors along a 15-m corridor following the current international recommendations.[35] The $SpO_2$ was monitored throughout the test with a digital pulse oximeter (Nonin Medical, Inc. Minneapolis, MN, USA).

### S100B measurement

Blood serum was obtained by centrifuging the blood samples for 10 min at 4,000 rpm and was kept frozen at −80°C until studied. Serum samples were analyzed for human S100B using commercial enzyme-linked immunosorbent assay kits (EMD Millipore, Billerica, MA, USA). S100B concentrations are expressed in pg/mL and the limit of detection was 2.74 pg/mL. More details on S100B measurement can be found in the supplemental material.

### Torque and electromyography recordings

Maximal quadriceps torque was studied during isometric maximal voluntary contractions (MVCs) of the dominant leg with hip and knee angles set at 90° and using the same settings as previously described.[2] The surface electromyography (EMG) signal of the vastus medialis was recorded using bipolar, silver chloride, surface electrodes. The surface EMG signal was amplified (×1000) and recorded at a sampling frequency of 4096 Hz (Biopac MP100, Biopac Systems, Santa Barbara, CA, USA).

### Neuromuscular excitability and activation

Peripheral nerve stimulation was used to measure peripheral voluntary activation (peripheral VA), muscle contractility (peak twitch), muscle excitability (M-wave), and spinal excitability (H-wave). The femoral nerve of the dominant leg was stimulated with a constant-current, high-voltage stimulator (DS7AH, Digitimer, Hertforshire, UK). A recruitment curve was performed at rest to determine which intensities to use during the protocol to elicit maximal M-waves (Mmax) and H-waves (Hmax).

Transcranial magnetic stimulation was used to measure cortical voluntary activation (cortical VA) and corticospinal excitability. Single transcranial magnetic stimulation (TMS) pulses of 1-ms duration were delivered over the motor cortex using a Magstim 200 (Magstim Co., Whitland, UK). A recruitment curve was performed during voluntary contraction at 10% of the maximal quadriceps torque in order to determine the maximal intensity.[36] The intensity at which the highest motor-evoked potentials (MEP) was observed was then used during the protocol to assess cortical VA and corticospinal excitability.

More details on the peripheral nerve and transcranial magnetic stimulation procedures are provided in the supplemental material.

### Protocol

The neuromuscular tests consisted of four MVCs of the knee extensors, each separated by 2 min of recovery (Figure 2). Participants were asked to maintain maximal effort for at least 4 sec. A double pulse at 100 Hz was delivered at the Mmax intensity over the femoral nerve during the force plateau of the first two MVCs (superimposed doublet) and 2 sec after relaxation (control doublet), according to the twitch interpolation technique.[37] A single transcranial magnetic stimulation pulse was delivered over the motor cortex to elicit MEPs during the force plateau of the last two MVCs. Three single pulses at Mmax intensity or Hmax intensity separated by 10 sec were delivered twice between MVCs to elicit Mmax and Hmax at rest, respectively (see Figure 2 for more details). After the MVCs, three submaximal voluntary contractions (SVCs) with visual feedback were performed at 50% and 30% of MVC. A single transcranial magnetic stimulation pulse was delivered during the force plateau of each SVC to elicit superimposed twitch responses at 30% and 50% of MVC. Then the transcranial magnetic stimulation resting twitch was determined by extrapolation of the linear regression between voluntary force and the superimposed twitch evoked at 30%, 50%, and during MVC.[38]

### Data Analysis

Maximal quadriceps torque ($Q_{MVC}$) was selected as the highest torque plateau of 500 ms from the four MVCs. Muscle contractile properties were evaluated by the quadriceps peak twitch ($Q_{Pt}$) from the highest twitch response induced by femoral nerve stimulation at rest.

Muscle excitability was determined as the highest Mmax peak-to-peak amplitude induced by femoral nerve stimulation at rest.

Spinal excitability was determined as the highest Hmax peak-to-peak amplitude normalized with respect to muscle excitability (i.e., Hmax/Mmax).

The amount of neural drive to the muscle was quantified by the root mean square of the vastus medialis EMG signal ($EMG_{RMS}$) during the highest torque plateau of 500 ms normalized with respect to muscle excitability (i.e., $EMG_{RMS}$/Mmax).

Peripheral VA was calculated via femoral nerve stimulation according to the twitch interpolation technique[37] as follows:

Peripheral VA (%) = [1 − ((superimposed doublet)∕(control doublet)) × 100]

Motor cortex activation (cortical VA) was calculated via transcranial magnetic stimulation. Because the relationship between superimposed transcranial magnetic stimulation twitch and voluntary force is not linear for intensities below 25% of MVC (reduced cortical and spinal excitability at low force levels[39]), the transcranial magnetic stimulation resting twitch was estimated by extrapolation of the linear regression between voluntary force and the superimposed twitch evoked at 30% of MVC, 50% of MVC, and during MVC.[38] The cortical VA was calculated as follows[38]:

Cortical VA (%) = [1 − ((superimposed twitch)∕(estimated resting twitch)) × 100]

Corticospinal excitability was assessed by the amplitude of the maximal MEP induced by transcranial magnetic stimulation during MVCs, normalized with respect to muscle excitability (i.e., MEP/Mmax). The cortical silent period (CSP) duration was measured as the time between MEP onset and the return of voluntary EMG activity. The central motor conduction time (CMCT) was calculated from the delay between the stimulus artifact and MEP onset.

### Statistical Analysis

All statistical analyses were performed using Statistica software (StatSoft, Inc., version 6.0, Tulsa, OK, USA). All data were examined for normality using a Shapiro-Wilk test. Differences between $NREM_{Des}$ and $NREM_{noDes}$ patients were studied using unpaired *t*-tests for parametric data, and nonparametric Mann-Whitney *U* tests otherwise. The required sample size for the substudy was calculated on the level of voluntary activation (main outcome), based on a between-groups difference of 20%.[40] With a 5% significance level and a power of 90%, the required sample size was ten per group. Data are reported as mean and standard deviation (SD) or median and quartiles (lower and upper quartiles labeled respectively by Q1 and Q3) in the case of nonparametric statistics. The significance level was set at P ≤ 0.05.

### RESULTS

### Prevalence of NREM Sleep Desaturation

The main characteristics of the $NREM_{noDes}$ and $NREM_{Des}$ patients are depicted in Table 1. The $NREM_{Des}$ group was composed of 59 patients (51.3% of the study sample), meaning that over half of the patients with COPD spent more than 10% of NREM sleep time with $SpO_2$ below 90%. Mean $SpO_2$ during NREM sleep was 92.9 ± 1.51% in the $NREM_{noDes}$ patients versus 88.9 ± 1.96% in the $NREM_{Des}$ patients (P < 0.001). There was no significant difference between the $NREM_{Des}$ and $NREM_{noDes}$ patients regarding age (P = 0.78), weight (P = 0.98), body mass index (BMI; P = 0.71), $FEV_1$ (P = 0.32), $FEV_1$/FVC (P = 0.13), blood gases (P = 0.15 and P = 0.98 for $PaO_2$ and $PaCO_2$, respectively) or AHI (P = 0.81).

### Subsample Characteristics and Blood Sample Analysis

The $NREM_{Des}$ and $NREM_{noDes}$ patients who took part in the neuromuscular tests (n = 29) did not exhibit any significant differences regarding age, weight, BMI, $FEV_1$, $FEV_1$/FVC, blood gases, or time to desaturate during exercise (Table 2). The total sleep time, arousal index, and AHI were also comparable between groups (P = 0.26, 0.97, and 0.92, respectively). Serum levels of S100B were significantly higher in the $NREM_{Des}$ compared with $NREM_{noDes}$ patients (P = 0.028). The values were 45.1 [Q1: 37.7, Q3: 62.8] versus 32.9 [Q1: 25.7, Q3: 39.5] pg.mL$^{-1}$ in the $NREM_{Des}$ and $NREM_{noDes}$ patients, respectively (Figure 3).

### Quadriceps Torque and Voluntary Activation

The data are presented in Figure 4. There were no significant differences on $Q_{MVC}$ (P = 0.58) or $Q_{Pt}$ (P = 0.48) between the $NREM_{Des}$ and $NREM_{noDes}$ patients. $Q_{MVC}$ values were 101.1 ± 39 and 110.9 ± 61 Nm, and $Q_{Pt}$ values were 41 ± 20 and 37 ± 16 Nm, for the $NREM_{Des}$ and $NREM_{noDes}$ patients, respectively. Conversely, peripheral VA was significantly lower in the $NREM_{Des}$ patients (90.7 ± 7.6 versus 95.9 ± 3.3%, P = 0.022). The cortical VA was also decreased in the $NREM_{Des}$ group compared with $NREM_{noDes}$ and was 89.5% [Q1: 85.8, Q3: 93.6] versus 94.1% [Q1: 93.6, Q3: 96.8], respectively (P = 0.03, Figure 4B).

### Electrophysiological Data

The data are presented in Table 3. $EMG_{RMS}$/Mmax and MEP/Mmax were significantly lower in the $NREM_{Des}$ compared

| Table 1—Characteristics of the patients included in the study. | | | | |
|---|---|---|---|---|
| | Total Sample | NREM$_{noDes}$ | NREM$_{Des}$ | P |
| n (% total sample) | 115 (100%) | 56 (48.7%) | 59 (51.3%) | |
| Sex M/F | 61/54 | 29/27 | 32/27 | |
| Age, y | 64.28 (9.2) | 64.04 (9.3) | 64.53 (9.1) | 0.78 |
| Weight, kg | 79.5 (19.2) | 79.5 (19) | 79.5 (19.6) | 0.98 |
| BMI, kg.m$^{-2}$ | 28.7 (6.31) | 28.5 (6.03) | 28.9 (6.6) | 0.71 |
| FEV$_1$, L | 1.48 (0.56) | 1.53 (0.58) | 1.43 (0.54) | 0.32 |
| FEV$_1$, % of predicted values | 55.7 (15.5) | 56.9 (15.5) | 54.6 (15.5) | 0.44 |
| FEV$_1$/FVC % | 54.1 (10.7) | 55.7 (10.4) | 52.6 (10.9) | 0.13 |
| PaO$_2$, mmHg | 68.7 (5.2) | 69.4 (5.78) | 68 (4.54) | 0.15 |
| PaCO$_2$, mmHg | 39.4 (5.56) | 39.5 (5.8) | 39.4 (4.5) | 0.98 |
| SaO$_2$, % | 92.9 (2.39) | 93.3 (2.41) | 92.5 (2.32) | 0.07 |
| AHI, events.h$^{-1}$ | 6.58 (4.93) | 6.43 (5.38) | 6.73 (4.5) | 0.81 |
| % of NREM sleep time with SpO$_2$ < 90% | 29 (35.4) | 0.89 [0, 2.1] | 46.2 [25.3, 89.9] | < 0.001 |

Values are means (standard deviation) or median [Q1, Q3 quartiles] in the case of nonparametric statistics. % of NREM sleep time with SpO$_2$ < 90% is the percentage of time spent with pulse oxygen saturation below 90% during the NREM sleep stage. AHI, apnea-hypopnea index; BMI, body mass index, FEV$_1$, forced expiratory volume in 1 sec, FVC, forced vital capacity, PaCO$_2$, arterial carbon dioxide tension; PaO$_2$, arterial oxygen tension; SaO$_2$, arterial oxygen saturation.

| Table 2—Characteristics of the patients who took part in the neuromuscular tests. | | | |
|---|---|---|---|
| | NREM$_{noDes}$ (n = 15) | NREM$_{Des}$ (n = 14) | P |
| Sex M/F | 10/5 | 9/5 | |
| Age, y | 61.5 (8.57) | 61.7 (6.09) | 0.93 |
| Weight, kg | 69.5 (18.3) | 74.6 (19.3) | 0.46 |
| BMI, kg.m$^{-2}$ | 25 (6.66) | 25.9 (5.85) | 0.69 |
| FEV$_1$, L | 1.28 (0.54) | 1.36 (0.57) | 0.71 |
| FEV$_1$, % of predicted values | 45.9 (15.5) | 49.1 (16.8) | 0.61 |
| FEV$_1$/FVC % | 46.5 (11.5) | 46.6 (11.5) | 0.98 |
| PaO$_2$ mmHg | 73.5 (6.33) | 71.6 (10.5) | 0.56 |
| PaCO$_2$ mmHg | 38.9 (3.78) | 41.1 (5.85) | 0.35 |
| SaO$_2$% | 94.4 (1.68) | 93.7 (3.04) | 0.44 |
| % of 6 MWT time with SpO$_2$ < 90% | 41.1 (35.6) | 59.5 (40.7) | 0.20 |
| Total sleep time, min | 370.2 (92.4) | 324.9 (86.8) | 0.26 |
| Arousal index, events.h$^{-1}$ | 16.1 (9.64) | 16.3 (11.66) | 0.97 |
| AHI, events.h$^{-1}$ | 6.77 (7.91) | 7.42 (6.81) | 0.84 |
| % of NREM sleep time with SpO$_2$ < 90% | 0.6 [0, 5] | 50.45 [16.6, 69] | < 0.001 |

Values are means (standard deviation) or median [Q1, Q3 quartiles] in the case of nonparametric statistics. % of 6 MWT time with SpO$_2$ < 90% is the percentage of time spent with pulse oxygen saturation below 90% during the 6 min walking test. % of NREM sleep time with SpO$_2$ < 90% is the percentage of time spent with pulse oxygen saturation below 90% during the nonrapid eye movement (NREM) sleep stage. AHI, apnea-hypopnea index; BMI, body mass index; FEV$_1$, force expiratory volume in 1 sec; FVC, forced vital capacity; PaCO$_2$, arterial carbon dioxide tension; PaO$_2$, arterial oxygen tension; SaO$_2$, arterial oxygen saturation.

with NREM$_{noDes}$ patients (P = 0.031 and P = 0.03, respectively). Mmax amplitude (P = 0.08), Hmax/Mmax (P = 0.66), CSP (P = 0.28) and CMCT (P = 0.88) were not significantly different between groups.

## DISCUSSION

The major findings of the study were that more than half of the nonhypoxemic COPD patients spent more than 10% of NREM sleep time with SpO$_2$ below 90%, and the nonhypoxemic COPD patients who spent more than 10% of NREM sleep time in desaturation had reduced motor cortex activation and excitability during maximal voluntary contractions and higher serum S100B concentrations.

### Prevalence of NREM Sleep Desaturation

The prevalence of O$_2$ desaturation during sleep is thought to be in the range of 38% to 70% in nonhypoxemic COPD patients, and our data are consistent with this range.[17–19] In the current study, we used a cutoff of only 10% of the NREM sleep time, with SpO$_2$ below 90% to diagnose NREM sleep desaturation. As NREM sleep desaturation has never been specifically assessed in COPD, our choice was dictated by the criteria used to diagnose desaturation during TST. Although no clear consensus exists, a percentage of the TST with SpO$_2$ below 90% is frequently cited in the literature.[41] We opted for a 10% criterion with SpO$_2$ below 90% as it has classically been used to study brain impairment.[33]

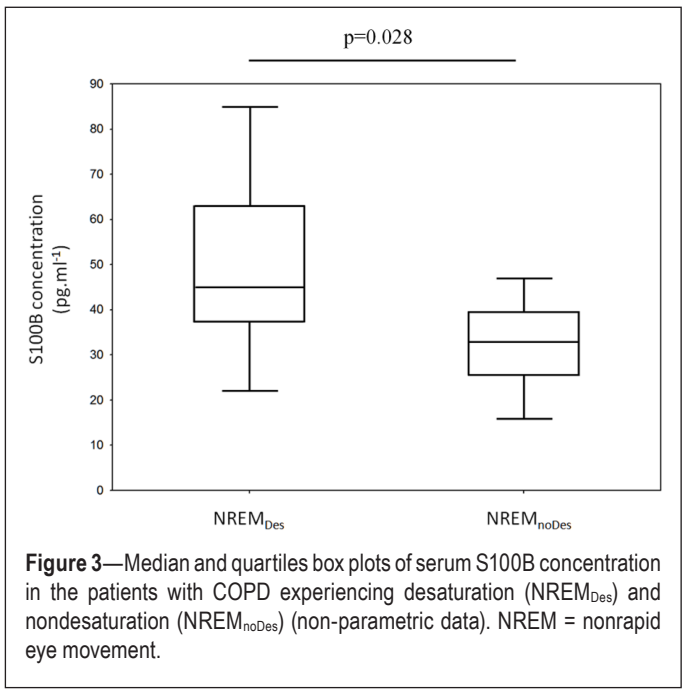

Figure 3—Median and quartiles box plots of serum S100B concentration in the patients with COPD experiencing desaturation (NREM$_{Des}$) and nondesaturation (NREM$_{noDes}$) (non-parametric data). NREM = nonrapid eye movement.

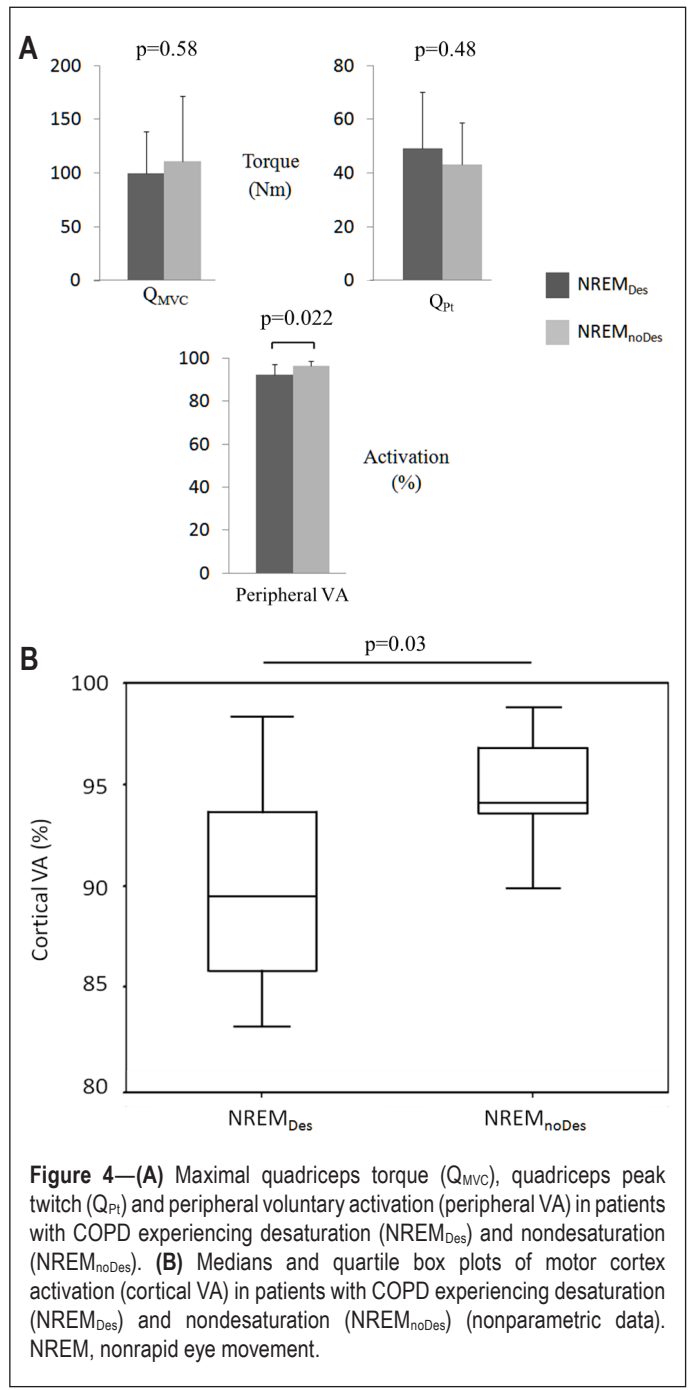

Figure 4—(A) Maximal quadriceps torque (Q$_{MVC}$), quadriceps peak twitch (Q$_{Pt}$) and peripheral voluntary activation (peripheral VA) in patients with COPD experiencing desaturation (NREM$_{Des}$) and nondesaturation (NREM$_{noDes}$). (B) Medians and quartile box plots of motor cortex activation (cortical VA) in patients with COPD experiencing desaturation (NREM$_{Des}$) and nondesaturation (NREM$_{noDes}$) (nonparametric data). NREM, nonrapid eye movement.

This study is the first to assess the prevalence of NREM sleep desaturation in COPD. Although it is indisputable that the deepest O$_2$ desaturation occurs during REM sleep,[41] a few studies have observed NREM sleep desaturation in patients with COPD who experience nocturnal desaturation.[42,43] Our results are consistent with these observations and provide further evidence that O$_2$ desaturation during sleep is not restricted to REM sleep in COPD.

It should be noted that the mechanisms responsible for NREM desaturation cannot be determined from our results, although the phenomenon seems unlikely to be explained by obstructive sleep apnea (OSA). Indeed, patients with severe OSA were excluded, and the AHIs were comparable in the two groups of patients with and without NREM sleep desaturation. Furthermore, the mechanisms are unlikely to involve the levels of diurnal PaO$_2$, given the absence of diurnal PaO$_2$ differences between the groups. This result is unsurprising because PaO$_2$ changes during sleep are not correlated with the diurnal PaO$_2$ levels in COPD.[44] Two important candidates to explain desaturation during sleep in COPD, especially during REM sleep, are alveolar hypoventilation and ventilation-perfusion mismatching.[41] Their potential implication in NREM sleep desaturation remains to be investigated.

### Effect of NREM Sleep Desaturation on Neuronal Damage and Neuromuscular Function

The second purpose of the study was to assess the repercussions of NREM sleep desaturation on neuronal damage and neuromuscular function. To do so, serum S100B, an easy-to-use and cost-effective biomarker of neuronal damage,[25–28] was analyzed in a subgroup of patients with COPD. The ability of serum S100B to detect brain impairment was recently confirmed in COPD and found to be associated with hippocampal atrophy and impaired cognitive function.[22] In the current study, we observed a higher S100B concentration in patients with NREM sleep desaturation, but without having the possibility to localize the impaired brain areas. The higher serum S100B concentrations could be linked to confounding factors other than NREM sleep desaturation, such as decreased sleep quality or sleep deprivation.[45] Importantly, we did not find any differences between the patients who did and did not experience desaturation during NREM sleep regarding TST and the arousal index.

| | NREM$_{noDes}$ (n = 15) | NREM$_{Des}$ (n = 14) | P |
|---|---|---|---|
| Mmax amplitude mV | 2.44 [1.64, 3.22] | 4.32 [2.12, 7.9] | 0.08 |
| Hmax/Mmax | 0.259 (0.208) | 0.220 (0.153) | 0.66 |
| EMG$_{RMS}$/Mmax | 0.077 (0.040) | 0.046 (0.029) | 0.031 |
| MEP/Mmax | 0.529 (0.143) | 0.285 (0.243) | 0.03 |
| CSP ms | 110 (20.3) | 101 (14.2) | 0.28 |
| CMCT ms | 20.6 (3.8) | 20.2 (5.71) | 0.88 |

Table 3—Electrophysiological responses to transcranial magnetic and femoral nerve stimulation.

Values are means (standard deviation) or median [Q1, Q3 quartiles] in the case of non-parametric statistics. CMCT, central motor conduction time; CSP, corticospinal silent period; EMG$_{RMS}$, root mean square of the vastus medialis electromyogram; MEP, motor-evoked potential; Mmax, maximal M-wave.

The functional repercussions of neuronal damage can be numerous. We chose to focus specifically on neuromuscular function and its effect on maximal quadriceps strength, as peripheral muscle weakness is one of the main deleterious systemic effects in COPD.[46] By stimulating the motor cortex, we observed lower MEP/M amplitude in the NREM$_{Des}$ patients during maximal voluntary contractions. The MEP/M amplitude reflects both spinal and cortical excitability.[47] In the current study, the comparable H-reflex amplitude observed in the NREM$_{Des}$ and NREM$_{noDes}$ patients indicates that the lower MEP/M could not be attributed to lower spinal excitability and thus is mainly explained by reduced motor cortex excitability. Furthermore, the reduced cortical excitability was associated with lower motor cortex activation as well as lower quadriceps motor unit activation (as measured by peripheral VA and EMG$_{RMS}$/M). These results support the hypothesis that patients with COPD who experience NREM sleep desaturation have reduced neural drive reaching the quadriceps muscle during MVC because of motor cortical output failure.

Impaired neural drive to the quadriceps has been a controversial topic in COPD. One study reported lower activation at the muscle level in patients with COPD compared with healthy controls,[40] whereas others did not.[48,49] More recently, we found lower cortical activity through neuroimaging assessment in patients with COPD during MVCs.[2] By using the neuroimaging technique, it was not possible to infer that the lower cortical activity resulted in lower cortical output.[2] In the current study, we assessed the cortical motor output with a more direct approach by stimulating the motor cortex. Our results confirm that cortical output is impaired in COPD but that it mainly concerns those patients with NREM sleep desaturation, as the values of voluntary activation reached by the NREM$_{noDes}$ patients (around 95%) were substantially similar to those of healthy subjects reported in other studies.[38,49] In addition, our data suggest that the discrepancies in previous studies might be explained by differences in the number of patients with COPD who experience NREM desaturation, which was not taken into account in previous works.

Maximal muscle torque depends in part on the ability to activate the muscle.[50] In addition, as the relationship between peripheral VA and torque is curvilinear, small modulations in peripheral VA induce much larger Q$_{MVC}$ changes at near-maximal contraction intensities.[51,52] For example, it was shown that a 5.7% increase in peripheral VA induced a 20.4% increase in Q$_{MVC}$.[52] Conversely, a 3% decrease in peripheral VA caused by neuromuscular fatigue has been associated with a 11% decrease in Q$_{MVC}$.[45] Therefore, the relatively low peripheral VA for NREM$_{Des}$ compared with NREM$_{noDes}$ patients (average of 5.2% less) should have expressed a greater loss of strength in these patients than the average of 9% (nonsignificant) strength reduction (101 versus 111 Nm, P = 0.58). The finding that the NREM$_{Des}$ patients reached the same torque level as the NREM$_{noDes}$ patients could be explained by low statistical power, or it may suggest a compensatory mechanism(s). Concerning the first explanation, it is important to note that the SD of the Q$_{MVC}$ data are in accordance with those of other studies.[53] In addition, the current Q$_{MVC}$ data are far from the level of statistical significance and we calculated the *a posteriori* number of subjects needed to obtain 90% statistical power (400 participants). These observations are in accordance with a limited experimental effect, if proven. Any potential compensatory mechanism is unlikely to be linked to a difference in intrinsic muscle capacity (due to higher muscle mass or contractility) because Q$_{Pt}$ was comparable between groups. Muscle torque at a joint is the result of contributions from both agonist and antagonist muscles. In a condition of decreased agonist torque (due to lower cortical activation), any lower torque developed by the antagonist knee flexor muscles during maximal quadriceps contraction in the NREM$_{Des}$ group could account for the comparable resultant torque; that is, comparable Q$_{MVC}$. Unfortunately, the antagonist activity was not assessed in this study, but this hypothesis is supported by a study carried out by Simoneau et al.[54] These authors reported no differences in the resultant maximal torque of the dorsiflexors in elderly subjects compared with young subjects, despite a 40% decrease in agonist maximal torque. This was explained by an activation of the antagonist plantar flexor muscle during maximal dorsiflexion that was almost twofold lower in the elderly, showing that in some circumstances maximal voluntary torque can apparently be preserved despite a significant decrease in agonist torque.

### Study Limitations

Serum S100B, which was used as a marker of CNS injury, has the advantage of being a strong, sensitive, and easy-to-use marker of neuronal damage.[29] However, although S100B is a marker of cerebral damage, it does not inform the location of the damage and cannot be used to localize the impaired brain areas. Computed tomography and MRI are likely to provide useful complementary information on the cerebral damage in patients with COPD experiencing desaturation during NREM sleep.

We did not directly assess the cerebrovascular $O_2$ reactivity during sleep. Therefore, although our results are highly consistent with an effect of NREM sleep desaturation on brain impairment, the occurrence of brain hypoxia during NREM sleep in the patients experiencing desaturation could only be inferred from the literature data.[12,13,55,56] A study to address the effect of correcting NREM sleep desaturation on serum S100B levels and motor cortex impairment would address this limitation.

## CONCLUSION

NREM sleep desaturation is far from negligible as it concerns approximately one of two patients with moderate to severe COPD and a resting $PaO_2$ between 60 and 80 mmHg. The patients with COPD who experience desaturation during NREM sleep exhibited an elevated level of a biomarker of CNS injury (i.e., serum S100B) and lower neural drive during quadriceps MVCs due to impaired cortical motor output. The observation that quadriceps muscle weakness was not more marked in the patients who experience desaturation suggests the existence of compensatory mechanisms whose nature and origin remain to be determined. Overall, the results are consistent with an involvement of NREM sleep desaturation in triggering CNS injury and decreasing neural drive to the quadriceps in COPD. The prevention of NREM sleep desaturation may well be an important clinical perspective to promote cerebral plasticity in COPD. Further studies are needed to determine the extent to which reversing neural activity is beneficial for the maximal voluntary force and functional capacity of patients with COPD.

**ACKNOWLEDGMENTS**

The authors thank Professor Robin Candau and Dr. Henri Bernardi for assistance and the use of their facilities for serum data analyses, and Dr. Mathieu Gueugnon for assistance in Matlab analyses. Furthermore, the authors also thank the patient's association Apard for the use of the polysomnograph.

**SUBMISSION & CORRESPONDENCE INFORMATION**

Submitted for publication January, 2015
Submitted in final revised form August, 2015
Accepted for publication September, 2015
Address correspondence to: Francois Alexandre, Movement To Health (M2H), Euromov, University of Montpellier, 700 avenue du Pic Saint Loup, 34090 Montpellier, France; Tel: (+33) 434 432 632; Fax: (+33) 434 432 644; Email: alexandre.francois88@gmail.com

**DISCLOSURE STATEMENT**

This was not an industry supported study. Dr. Alexandre was partially supported by a grant in aid from the French Ministry through the "Association Nationale de la Recherche et de la Technologie" (National Agency for Research and Technology). The authors have no other funding to declare. The authors have indicated no financial conflicts of interest. The work was performed at the Clinique du Souffle La Vallonie, Fontalvie, 800 avenue Joseph Vallot, 34700 Lodève, France, and Clinique du Souffle Les Clarines, Fontalvie, Route de Condat, Le Sédour Sud, 15400 Riom-es-Montagnes, France.